\documentclass[structabstract]{aa}

\usepackage{natbib}
\bibpunct{(}{)}{;}{a}{}{,} 
\usepackage{graphicx}
\usepackage{txfonts}
\usepackage{ulem}
\usepackage{color}

\newcommand{\xmm}{{{XMM-Newton}}}
\newcommand{\rosat}{{{ROSAT}}}
\newcommand{\txmm}{{\fontfamily{ptm}\fontseries{m}\fontshape{sc}\selectfont{3XMM}}}
\newcommand{\omc}{{\fontfamily{ptm}\fontseries{m}\fontshape{sc}\selectfont{OMC}}}

\begin{document}

\title{AGN spectral states from simultaneous UV and X-ray observations by XMM-Newton}

\author{J.~Svoboda,$\!^{1}$, M.~Guainazzi,$\!^{2,3,4}$ A.~Merloni$\,^{5}$ }
\institute{$^{1}$ Astronomical Institute, Academy of Sciences, Bo\v{c}n\'{\i}~II~1401, CZ-14131~Prague, Czech~Republic,\\
$^{2}$ European Space Astronomy Centre of ESA, P.O. Box 78, Villanueva de la Ca\~{n}ada, 28691 Madrid, Spain,\\ 
$^{3}$ European Space Research \& Technology Centre, Postbus 299, 2200 AG Noordwijk, Netherlands,\\ 
$^{4}$ Institute of Space and Astronatical Science (JAXA), 3-1-1 Yoshinodai, Sagamihara, Kanagawa, 252-5252, Japan,\\
$^{5}$ Max-Planck-Institut f\"{u}r extraterrestrische Physik, Giessenbachstrasse 1, 85748 Garching, Germany
}
%

\authorrunning{J.~Svoboda et al.}
\titlerunning{AGN-GBHC states}
\offprints{J.~Svoboda, email: jiri.svoboda@asu.cas.cz}

\abstract
{
It is generally believed that the supermassive black holes in active galactic nuclei (AGN) 
and stellar-mass black holes in X-ray binaries (XRBs) work in a similar way. 
}
{
While XRBs evolve rapidly and several sources have undergone a few complete cycles from  quiescence to an outburst and back, 
most AGN remain in the same state over periods of years and decades, due to their longer characteristic  timescale proportional to their size. 
However, the study of the AGN spectral states is still possible with a large sample of sources. 
Multi-wavelength observations are needed for this purpose since the AGN thermal disc emission 
dominates in the ultraviolet energy range, while the up-scattered hot-corona emission is detected in X-rays.
}
{
 We compared simultaneous UV and X-ray measurements of AGN obtained by the XMM-Newton satellite. 
 The non-thermal power-law flux was constrained from the 2-12\,keV X-ray luminosity,
 while the thermal disc component was estimated from the UV flux at $\approx 2900 \AA$. 
 The hardness (defined as a ratio between the X-ray and UV plus X-ray luminosity)
 and the total luminosity were used to construct the AGN state diagrams.
 For sources with reliable mass measurements, the Eddington ratio was used instead of the total luminosity.
}
{
The state diagrams show that the radio-loud sources have on average higher hardness,
due to the lack of the thermal disc emission in the UV band,
and have flatter intrinsic X-ray spectra.
In contrast, the sources with high luminosity and low hardness
are radio-quiet AGN with the UV spectrum consistent with the multi-temperature
thermal disc emission.
The hardness-Eddington ratio diagram 
reveals that the average radio-loudness is stronger for low-accreting sources,
while it decreases when the accretion rate is close to the Eddington limit.
}
{
Our results indicate that the general properties of AGN accretion states are similar 
to those of X-ray binaries. 
This suggests that the AGN radio dichotomy of radio-loud and radio-quiet sources 
can be explained by the evolution of the accretion states.
}



\maketitle

\section{Introduction}\label{intro}

Astronomical observations have confirmed two classes of black holes:  stellar-mass black holes 
in X-ray Binaries (XRBs) and supermassive black holes in active galactic nuclei (AGN). 
The black hole mass ranges from a few solar masses in XRBs to millions and billions 
of solar masses in AGN, which implies different sizes and timescales
(for a recent review on AGN,  see e.g.
\citealt{Merloni2016, Netzer2015} and references therein;
for XRB, see e.g. \citealt{2006ARA&A..44...49R, 2007A&ARv..15....1D, Fender2012, Zhang2013}
and references therein).
 
A typical timescale of a complete cycle of an XRB outburst from quiescence
through hard state to soft state is of the order of a few hundred days \citep{2006ARA&A..44...49R, Dunn2010}.
This translates to $10^5 - 10^9$ years for AGN depending on the black hole mass.
The shortest independently estimated timescales for an AGN phase 
has been reported by \citet{Schawinski2015} to be $10^5$ years.
It is therefore quite unlikely, though not impossible, to detect a spectral state
transition in an AGN. Some changing-look AGN were proposed to be candidates
of such spectral changes \citep{LaMassa2015, MacLeod2016, McElroy2016}, 
but in some cases other interpretations have also been proposed \citep[see e.g.][]{Merloni2015}.
Most AGN, however, are not variable in X-rays by a factor larger than a few over decades
\citep{Strotjohann2016},
suggesting that they remain in the same spectral state.

Nevertheless, accretion on the supermassive black holes is believed to be similar 
to accretion on the stellar-mass black holes \citep[see e.g.][]{Merloni2003, Koerding2006}.
X-ray binaries are known for their specific evolutionary track
in a hardness vs. intensity (or luminosity) diagram (`q-shaped' or `turtle-head').
The states evolve from the low-hard state up to the high state (still hard), 
then turning to the high-soft state
with their flux dominated by the thermal emission.
In the transition to the soft state,
sources are observed to cease the production of strong radio emission.
This is explained by quenching of the jet, and therefore 
this transition is represented by a jet line in the hardness intensity diagram \citep{Fender2004}.
Only temporary blob ejections, also known  as ballistic jets \citep{Narayan2012},
are reported in the transition from the hard to high-soft state, 
when sources occasionally cross back over the jet line in small cycles
(these states are called `steep power-law' or `intermediate' states; \citealt{Fender2004}).
Finally, the luminosity decreases and the source
continuously declines to the low-hard state.


\citet{Koerding2006a} and \citet{Sobolewska2011} have suggested that 
different types of active galaxies correspond to different spectral states of 
XRBs. Bright Seyfert galaxies are supposed 
to correspond to thermal soft states, where an optically thick and geometrically thin disc is assumed 
to extend down to the innermost stable circular orbit around a black hole.
The XRB spectrum in the soft state is dominated by the thermal emission of the accretion disc. 
The thermal emission shifts to the UV band for AGN \citep{Malkan1982, Laor1989}.
However, it was  noticed by \citet{Elvis1994} that the quasar spectra
cannot be simply described by the thermal black-body emission of a thin accretion disc.
Also, X-ray spectra are different  to X-ray binaries.
AGN seem to have an additional spectral component
that appears between the UV and X-ray bands, 
 which can be explained by additional  Comptonisation \citep{Done2012, Done2014a}.

At the low-luminosity regime, a specific class of AGN with low-ionised nuclear emission-line regions (LINERs) 
seems to  correspond well to the low-hard states of XRBs \citep{Sobolewska2011}.
In the low-hard state, the innermost regions of the disc become optically thin 
and the accretion flow is advection-dominated \citep{1997ApJ...489..865E}.
In this state, non-thermal processes prevail in the spectrum 
and the characteristic spectral slope is significantly harder. 
The analogy between AGN and XRBs is also seen in their timing properties \citep{McHardy2006}. 

\citet{Koerding2006a} have also suggested that different spectral states may explain the AGN radio-dichotomy. 
Most of AGN are radio-quiet while about 10-20\% are radio-loud \citep{Kellermann1989}. 
The radio emission is due to a relativistic jet launched close to the supermassive black hole. 
However, it is not clear whether the AGN powerful jets are related to the most rapidly rotating black holes 
with the energy released by the Blandford-Znajek mechanism \citep{1977MNRAS.179..433B}
or whether they are non-persistent features depending on the current accretion rate onto a supermassive black hole. 
In XRBs, the jet is not present during the thermal state \citep[see e.g.][and references therein]{2009MNRAS.396.1370F}.

The radio-loud AGN could therefore correspond to the hard states
\citep{Falcke2004,Koerding2006a}. 
\citet{Merloni2008} pointed out that two different states can be distinguished
among radio-loud sources. While radio-loud low-luminosity AGN 
(low kinetic mode) correspond to the low-hard states of X-ray binaries,
radio-loud quasars at high luminosity
(high kinetic mode) can be  
more likely
associated with the giant radio flares seen in XRB transition
(the `jet line' of \citealt{Fender2004}).

All previous studies on AGN spectral states are based on UV and X-ray fluxes measured non-simultaneously. 
\citet{Koerding2006a} have generated a disc-fraction luminosity diagram for a large sample
of quasars from the {{Sloan Digital Sky Survey}} \citep{2006ApJS..162...38A}
and from archival X-ray measurements from the {{ROSAT All-Sky Survey}} \citep{1999A&A...349..389V}.
They show that radio-loud AGN lack the thermal emission and their luminosity is dominated by X-rays from non-thermal processes.
However, their results can be affected by the limited bandpass of the ROSAT/PSPC detector,
and by the non-simultaneity of the data if the observed flux is significantly variable. 

The X-ray variability is caused either by the changes in the 
intrinsic X-ray emission or by variable obscuration of the circumnuclear gas. 
Depending on the absorber’s location, the characteristic timescale of the latter can be from months to years if the absorber 
is part of a distant torus \citep{2002ApJ...571..234R, Miniutti2014}, or from days to weeks if the absorber is located closer in the so-called broad-line region,
 e.g. NGC 4388 \citep{2004ApJ...615L..25E}, 
NGC 1365 \citep{2005ApJ...623L..93R}, NGC 4151 \citep{2007MNRAS.377..607P}, 
NGC 7582 \citep{2009ApJ...695..781B}, 
PG1535+547 \citep{2008A&A...483..137B}, 
and Fairall~51 \citep{Svoboda2015}.

In this paper, we aim to extend the previous analysis by \citet{Koerding2006a} 
in five very important aspects:\\
1) only simultaneous UV and X-ray data are considered;\\
2) 2-12\,keV X-ray flux by {\xmm} are used, and {\rosat} operated in 0.1-2.4\,keV
energy range;\\
3) UV flux measurements are used to estimate accretion-disc luminosity instead of an approximation from the optical range;\\
4) spectral slopes in UV and X-ray energy domains are used as an indication of the accretion state;\\
5) deeper observations are used, resulting in a sample that has a lower sensitivity limit.\\

The first two aspects are motivated mainly by the large X-ray variability
previously reported for several AGN. 
The simultaneity and the extended X-ray energy range should 
minimise the effect of variable X-ray absorption on the source luminosity estimates. 
The 2-12\,keV energy band also provides better estimates of the energetics and total X-ray luminosity,
since the power emitted by the corona emerges primarily in the hard X-rays.
The last two improvements are achievable thanks to the higher sensitivity
of the current instruments (on board XMM-Newton).

The paper is organised as follows. Section 2 describes which data resources
were employed when creating the sample.
Section 3 shows the hardness-intensity diagrams for AGN 
based on simultaneous measurements by the XMM-Newton satellite. 
The results are discussed in Section 4. 
The main conclusions are summarised in Section 5.


\section{Method}\label{method}


%
%
%


\begin{table}[tb!]

\caption{Main sample.}
\centering
\begin{tabular}{cc}
%
\rule{0cm}{0.5cm}
Flux and spectral information & Number of sources \\
\hline \hline
\rule[-0.7em]{0pt}{2em} 
simultaneous UV and X-ray flux & 1522 \\
\rule[-0.7em]{0pt}{2em} 
+ radio flux measurement (upper limit) & 175 (680) \\
\end{tabular}

\label{final_sample} 
\end{table}

\subsection{Data resources and sample creation}
\label{samples}

We used publicly available archives of X-ray and UV measurements performed by the XMM-Newton satellite
 in the years 2000-2015. The most recent X-ray catalogue is the third XMM-Newton
serendipitous source catalogue \citep{Rosen2016} (hereafter {\txmm}).
The UV data are taken from the XMM-Newton serendipitous ultraviolet source survey \citep{page2012},
version 2.1 (hereafter {\omc}). 


The UV and X-ray catalogues were cross-correlated 
to ensure that the measurements in the two bands are quasi-simultaneous.
The match was found to be positive because, 
first, the identification numbers of the observations were identical
and, second, the coordinates of the UV and X-ray detections
were within 3 arcsec, which is the nominal accuracy of the astrometric reconstruction. 
We note that the identity of the identification numbers does not ensure a strict simultaneity.
However, the difference in start/end times of the UV and X-ray exposures ($\approx$ minutes)
is negligible with respect to the typical variability timescales in AGN.

To select only AGN with measured redshift, 
we  cross-correlated this catalogue with the AGN catalogue
by \citet{Veron-Cetty2010}, 
and with the catalogue of quasars from the SDSS survey,  data release 12 \citep{Alam2015}.
We also added AGN from the SDSS (DR12) archive 
that were classified as galaxies of the subclass `AGN'
(using CasJobs\footnote{skyserver.sdss.org/casjobs/} for the search query).
Finally, we also used sources from the XMM-COSMOS survey
\citep{Hasinger2007, Scoville2007, Lusso2012}.
For all cross-matches with the {\xmm} catalogue, we used 3\,arcsec
as the maximum distance between the coordinates from the two catalogues.

Using the Veron catalogue, we also
excluded all sources classified as nuclear HII regions. These are the sources
that were originally considered as active galaxies, but were re-classified later.
Their activity is not primarily from accretion onto the central supermassive black hole,
but rather from a large star formation in the nuclear region.
We also excluded known blazars from our sample. 
Their X-ray flux is dominated by a strongly beamed jet emission, 
and therefore they cannot be used to study the relation between the thermal emission of the accretion disc and the non-thermal contribution from the corona,
the main astrophysical goal of our study.

We also removed sources 
with low significance of the UV detection
(the critical UV significance is 3\,$\sigma$)
and with extended UV emission 
(FWHM major-axes greater than the calibrated PSF FWHM with $>$ 3\,$\sigma$ confidence).
The latter is to eliminate uncertainties
when the UV emission might not correspond to the UV emission from the accretion disc.
We did not consider underexposed  observations as well.
An observation is considered  underexposed
when the exposure time in either the UV or the X-rays is lower than 1\,ks 
or the uncertainty in an UV or X-ray flux measurement exceeds 100\%.
Furthermore, we also removed sources with flat ($\Gamma < 1.5$) or steep ($\Gamma > 3.5$)  X-ray slopes
(see Sect.~\ref{powerlaw_luminosity}, determining  the X-ray slope
from the XMM-Newton catalogue).

Finally, we  removed sources for which the only available UV/optical measurements
correspond to the UV flux at lower wavelengths than 1240{\AA} in their rest frame.
This flux could be contaminated by Ly$\alpha$ emission.
Moreover, even if the thermal emission  dominated,
it might already correspond to a high-energy tail
instead of the increasing part (in the energy domain)
heading towards the thermal peak. Since the position of the thermal peak
depends on a generally unknown combination of black hole mass and spin,
we cannot make any predictions about the UV slope in such cases,
which is important for the extrapolation of the UV flux to a reference wavelength 
(see Sect.~\ref{disc_luminosity}).

If a source was observed multiple times by {\xmm}, we retained in our sample only
the observation corresponding to the best combination of the X-ray exposure time
and the UV significance. 
We defined a total observation significance as the sum 
of a ratio between the exposure and the maximum exposure of the source 
and a ratio between the UV significance and the maximum UV significance of the source.
The observation with the highest total significance was selected,
so that each source is represented by one measurement only.
This is to have a homogeneous representation of different sources,
which would allow us to compare number densities at different parts of the diagram 
with the same quantity for X-ray binaries.

\begin{figure}[tb!]
 \includegraphics[width=0.5\textwidth]{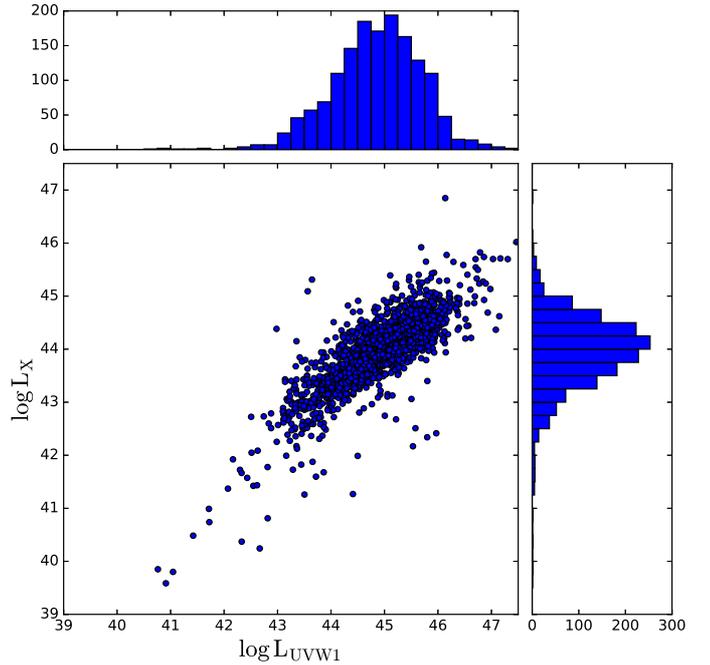}
\caption{Distribution of the measured UV and X-ray luminosities of our main sample.
Both quantities are also plotted in histograms on the sides.
Both luminosities are redshift corrected (see the main text for the details).}
\label{histogram_active}
\end{figure}

We refer to this created sample of sources as the main sample. It contains 1522 sources,
from which 680 sources have been observed in radio surveys.
However, some of them are only flagged as observed but with not constrained radio flux
in the V\'{e}ron catalogue. Only 175 sources have constrained radio-flux measurements.
They are mainly adopted from the 
VLA NVSS \citep{Condon1998} and FIRST \citep{Becker1995} surveys. 
Radio measurements at 20\,cm are preferentially taken from 
the cross-matched values of the SDSS and FIRST survey \citep{Becker1995}
from the SDSS archive, but when the radio measurements at 20\,cm 
were positive only in the Veron catalogue, we chose those values.
Table~\ref{final_sample} summarises the number of sources
for which radio measurements are available.
Further details on the main sample construction is given
in  Appendix~A1.

For further analysis, we use this set of cosmological parameters in our analysis:
$H_0 = 67.8$\,km\,s$^{-1}$\,Mpc$^{-1}$, $\Omega_M=0.308$, $\Lambda_0=0.69$ \citep{PlanckCollaboration2015}.
The analysis was done using the  Interactive Data Language (IDL) software, version 8.2.
The plots are created with Python using the Matplotlib and Numpy packages,
and also Gnuplot software.

\subsection{Correction on Galactic extinction and redshift}

The UV flux measurements are affected by the Galactic extinction.
For the de-reddening, we used a relation by \citet{Guver2009}:
\begin{equation}
\label{Galactic_extinction}
 E(B-V) = (1/0.22)*N_{\rm H}/R_V,
\end{equation}
where $N_{\rm H}$ is the column density taken from the LAB survey \citep{Kalberla2005},
and $R_V = 3.1$ specifies the ratio of the total to selective extinction \citep[see e.g.][]{Savage1979}.
We used an IDL procedure FM\_UNRED based on parametrisation by \citet{Fitzpatrick1999}.

Our sample is very heterogeneous in redshift.
It includes local AGN,
as well as more distant ones with redshift going up to z $\approx$ 3.
Redshift correction (also known as K-correction) of the observed fluxes is therefore inevitable.
For this purpose, we approximated the spectral flux density
as a power law in each energy range:
\begin{equation}
\label{nu}
 \nu F_{i} = \nu^{-\alpha_{i}},
\end{equation}
where $i = r, uv, x$ denotes  radio, UV, and X-ray energy domains, respectively.
The K-correction can be then prescribed as
\begin{equation}
 F_{\rm intrinsic} = F_{\rm observed} * (1+z)^{\alpha-1}
 \label{K-correction}
.\end{equation}

For the radio spectral slope, we adopted the value $\alpha_r = 0.5$ of an average AGN 
spectral energy distribution from the literature \citep{Ho2001}. 
The UV spectral energy distribution, however, has a very large dispersion 
among all AGN \citep{Richards2006}.
Therefore, for the observations where more UV filters had significant detection,
we constrained the UV slope directly from our data.
Since our sample spans a wide range in the redshift,
we always used the nearest UV or optical filters
to the observed-frame reference wavelength $(1+z)*\lambda_{\rm ref}$,
where $\lambda_{\rm ref}=2910\,\AA$.\footnote{We never used the UV flux measurements corresponding to the rest-frame
wavelength shorter than $\lambda_{\rm crit} = 1240\,\AA$, which is explained
in greater detail in Sect.~\ref{disc_luminosity}.}
The reference wavelength is chosen to correspond to the UVW1 band,
so that little or no  extrapolation is needed for sources at low redshift that are observed in this band.
We calculated the UV slope in the wavelength domain as
\begin{equation}
\label{beta}
 \beta = \frac{\log {F_{\rm a}}/{F_{\rm b}}}{\log{\lambda_{\rm a}}/{\lambda_{\rm b}}},
\end{equation}
where $F_{\rm a}$ and ${F_{\rm b}}$ are observed flux densities in the nearest filters to the  
reference wavelength in the observed frame,
and $\lambda_{\rm a}$, and $\lambda_{\rm b}$ are the mean wavelengths of the employed filters.
When only a single filter was available, we used $\beta_{\rm default} = -1.5$
based on previous UV studies of quasars \citep{Scott2004, Richards2006}.
We measured a mean value of our sample (when the UV slope could be constrained using eq.~\ref{beta}),
and we got $\beta_{\rm mean} \approx -1.4$ with a relatively large dispersion $\sigma \approx 1.4$,
which is consistent with previously reported measurements.
Since some measured UV slopes were very discrepant from this value,
we used $\beta_{\rm default}$ for K-corrections
of all sources with the measured UV slope outside the interval ($\beta_{\rm mean}-\sigma$, $\beta_{\rm mean}+\sigma$)
in order not to introduce any artificial increase/decrease of the source UV luminosity
by extrapolation using an unreliable power-law index.

Measured UV fluxes and the UV slope can be significantly affected
by the interstellar scattering and absorption in the host galaxy (intrinsic extinction).
We have not a priori included any correction for the intrinsic extinction
because the exact shape of the extinction curve is precisely known only for some nearby galaxies
and is dependent on the AGN spectral energy distribution,
metallicities, and other individual properties of galaxies \citep[see e.g.][]{Calzetti1994, Czerny2004, Gaskell2004}.

However, we do not expect a large effect
of the intrinsic extinction on constraining the disc luminosity.
The disc luminosity is estimated from the UV flux measurements at the 
reference wavelength in the rest frame of each galaxy.
Thus, all sources would be affected in a similar way if the extinction
curve is not dramatically different from one source to another. 
Systematic uncertainties due to the intrinsic extinction will be included
in an arbitrary conversion factor from the UV to the disc luminosity
(see the definition in Sect.~\ref{disc_luminosity}).
However, the intrinsic extinction can still affect
the measured UV slopes especially for low-luminosity sources, 
which we discuss in Sect.~\ref{discussion_extinction}.


The distribution of the K-corrected UV and X-ray luminosities of the main sample
is shown in Figure~\ref{histogram_active}.
The UV flux at the wavelength of the UVW1 filter, $L_{\rm UVW1}$, peaks
around $\log L_{\rm UVW1} \approx 45$,
while the X-ray luminosity, $L_{\rm X}$, is typically one magnitude lower,
$\log L_{\rm X} \approx 44$. In the next sections, we estimate the corona (power-law)
luminosity as an extrapolation of the X-ray luminosity
and the accretion disc luminosity as proportional to the measured UV luminosity.


\subsection{ Corona (power-law) luminosity}
\label{powerlaw_luminosity}

The corona (power-law) luminosity can be defined from the X-ray luminosity by extrapolating it
to the energy interval 0.1-100\,keV. Beyond these energy limits, the corona emission
is usually diminished by low- and high-energy cut-offs. For the power-law luminosity,
we can therefore write
\begin{equation}
\label{lp}
 L_P =  4\pi D_L^2\,F_{0.1-100\,\rm{keV}},
\end{equation}
where $D_L$ is the luminosity distance constrained from the redshift measurement $z$,
and $F_{0.1-100\,\rm{keV}}$
is the X-ray flux at the 0.1--100\,keV energy calculated by an extrapolation
of the observed 2--12\,keV flux
\begin{equation}
F_{0.1-100\,\rm{keV}} = F_{2-12\,\rm{keV}}
\frac{100^{-\Gamma+2}-0.1^{-\Gamma+2}}{12^{-\Gamma+2}-2^{-\Gamma+2}},
\end{equation}
where $\Gamma$ is the photon index.
The 100\,keV threshold\footnote{The uncertainty of the total X-ray flux 
is of the order of 10\% assuming the photon index $\Gamma = 1.7$
and the high-energy cut-off that varies between 60 and 300 keV.} 
is selected to roughly correspond to an average value of the high-energy cut-off
based on deep hard X-ray observations of type-1 AGN with Integral, Swift, and Nustar
\citep{Malizia2014, Fabian2015}.

The photon index $\Gamma$ can be constrained
from the flux measurements in two neighbouring X-ray bands in the {\txmm} catalogue.
When calculating the photon index $\Gamma$, we assumed that the X-ray flux
between energies E$_a$ and E$_b$ can be described as a power law:
\begin{equation}
  F_{\rm a-b} \propto \int_{E_a}^{E_b} \! E^{-\Gamma+1} dE. 
\end{equation}
Using $F_{\rm 0.5-2\,keV}$ and $F_{\rm 2-12\,keV}$ we get for the photon index
\begin{equation}
\Gamma \approx 2 + 3.144*{\log{\frac{F_{\rm 0.5-2\,keV}}{F_{\rm 2-12\,keV}}}}.
\end{equation}

We used the photon index as an indicator of a nature of the X-ray spectrum.
While  X-ray spectra that are too flat suggest a large effect of X-ray absorption,
 spectra that are too steep might be incompatible with AGN emission or suggest a low quality
of the hard-band X-ray spectrum.
Therefore, only sources with a photon index $\Gamma$ in the interval $1.5 < \Gamma < 3.5$
were considered in the sample creation. 
The mean value of the photon index of the main sample is $\Gamma_{\rm mean} = 1.7 \pm 0.3$.\footnote{
We also constrained the hard X-ray photon index from the energy intervals 
$2-4.5$\,keV and $4.5-12$\,keV,
as $\Gamma_{\rm hard} \approx 2 + 2.268*{\log{\frac{F_{\rm 2-4.5\,keV}}{F_{\rm 4.5-12\,keV}}}}$,
which has the advantage of avoiding the contamination by the soft X-ray absorption.
Its average value is consistent with the broad-band value $\Gamma_{\rm{hard}} = 1.7 \pm 0.8$,
but has a much larger dispersion, indicating a larger scatter,
which might be due to a lower signal-to-noise ratio in the narrower hard bands.
Therefore, we used  only the broad-band value of the X-ray photon index
as defined from the $0.5-2$\,keV and $2-12$\,keV energy bands.}
This value is consistent with reported values of the average photon index
of unobscured AGN in the literature \citep{Nandra1994, Steffen2006, Young2009, Liu2016}.

\subsection{ Thermal disc luminosity}
\label{disc_luminosity}

The thermal disc emission is characterised by a multi-colour black-body emission
from accretion rings \citep{1984PASJ...36..741M}, since the temperature decreases with the radius $r$,
\begin{equation}
 T = \left[{\frac{3c^6}{4 \pi G^2 \sigma}\left({\frac{r}{r_g}}\right)^{-3}\frac{\dot{M}}{M^2}}\right]^{\frac{1}{4}},
\end{equation}
where $r_g = \frac{GM}{c^2}$ is the gravitational radius, $M$ is the black hole mass,
$\dot{M}$ is the accretion rate, $\sigma$ is the Stefan-Boltzmann constant, $G$ is the gravitational constant, and $c$ is the speed of light.
The temperature at a given $r$ in terms of $r_g$
is then inversely proportional to the one-fourth power of the black hole mass (assuming $\dot{M} \propto M$).
\begin{equation}
\label{temperature}
 T_{r[r_g]} \propto M^{-\frac{1}{4}}.
\end{equation}
The characteristic peak temperature of the thermal emission is of the order of 1\,keV (3\,keV)
for a non-rotating (highly rotating) stellar-mass black hole.
This corresponds to 221\,\AA\ (73.6\,\AA) for a $10^6 M_\odot$,
and 1240\,\AA\ (413\,\AA) for a $10^9 M_\odot$ supermassive black hole.

For black hole mass less than $10^9 M_\odot$ and spin $a > 0$,
the rest-frame emission at longer wavelengths than 1240\,\AA\ always correspond
to the rising part (in the energy domain) of the thermal disc emission
and can be roughly approximated as a power law 
(therefore, we can use eq.~\ref{nu}).
However, at wavelengths shorter then 1240\,\AA\, the emission might instead correspond to the high-energy
tail of the thermal emission, and using a power-law approximation
of the UV energy distribution would be invalid.
Therefore, any measurements corresponding to rest-frame wavelengths
shorter than 1240\,\AA\ were not considered.

For all sources of the main sample, we constrained the UV flux at the rest wavelength $\lambda_{\rm ref} = 2910\,\AA$, 
which is the mean wavelength of the UVW1 band.
Owing to the cosmological redshift, different UV/optical filters might be closer to this rest-frame reference wavelength.
The procedure to constrain the UV flux at the rest-frame reference wavelength was the following.
First, we determined the rest-frame wavelengths of the mean wavelengths of all UV and optical filters
on the Optical Monitor, UVW2 (2120\,\AA\ at z=0), UVM2 (2310\,\AA\ at z=0), UVW1 (2910\,\AA\ at z=0),
U (3440\,\AA\ at z=0), B (4500\,\AA\ at z=0), and V (5430\,\AA\ at z=0).
Then, we chose the two nearest filters to the rest-frame wavelength $\lambda_{\rm ref} = 2910\,\AA$,
from which we constrained the UV spectral slope and the observed flux
at the rest-frame wavelength.
Finally, we multiplied the observed UV flux by a K-correction factor (1+z) to get
the intrinsic UV flux at the rest wavelength $\lambda_{\rm ref}$.

We used the redshift and Galactic extinction corrected UV flux
to estimate the disc luminosity $L_D$, which can be defined as
\begin{equation}
\label{ld}
L_D =  A * 4\pi D_L^2 \lambda F_{\lambda,\,2910\,\AA},
\end{equation}
where $D_L$ is the luminosity distance constrained from the redshift measurement $z$
and $A$ is an arbitrary factor, which can be chosen so that
the sum of the disc and power-law luminosity, $L_{\rm tot}$, roughly corresponds to the bolometric luminosity.
 
Therefore, we cross-correlated our sample with a sample of active galaxies
studied by \citet{Vasudevan2009}
who calculated the bolometric corrections from the simultaneous X-ray/UV and optical
measurements by XMM-Newton. We obtained 
13 sources present in both samples.\footnote{3C390.3, Mrk110, PG0953+414, PG1226+023, FAIRALL9, 3C120,     
Mrk1095, Mrk590, Mrk79, NGC5548, NGC7469, PG0052+251, PG2130+099}
The multiplicative factor was constrained 
using the MPFIT procedure \citep{Markwardt2009} 
as $A \approx 2.4$, 
with the standard deviation $\sigma_A \approx 0.6$.

\begin{figure}[tb!]
\includegraphics[width=0.49\textwidth]{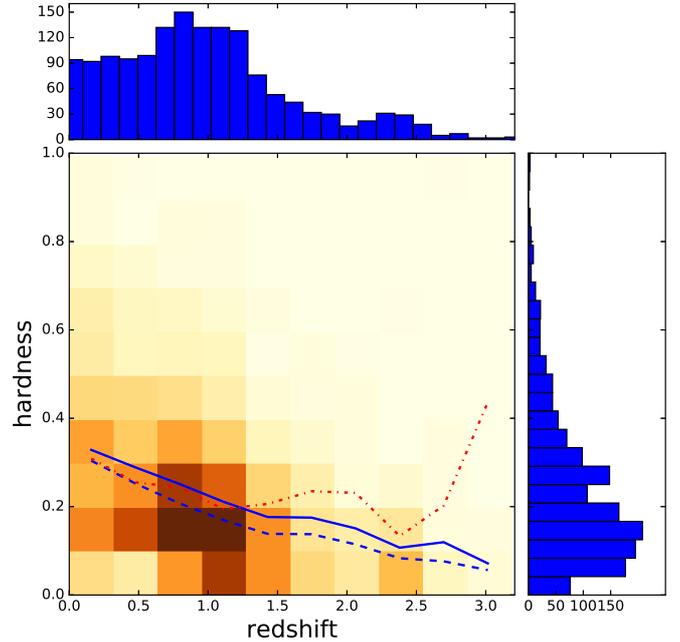}
\caption{Redshift-hardness distribution of the main sample.
The coloured 2D histogram shows the number of sources in a particular
bin of redshift and hardness interval. The inserted curves
indicate the mean value (blue solid line), median value (blue dashed line),
and the ratio of the median to mode (red dot-dashed line)  
of the hardness per  redshift bin. 
The side-bar histograms show the total number of sources
per hardness and redshift bins, respectively.}
\label{redshift}
\end{figure}


\subsection{Spectral hardness}

The diagram showing the spectral hardness on one axis and the X-ray intensity on the other
(the `Hardness-Intensity Diagram')
is widely used to track the spectral state evolution in black hole X-ray binaries \citep{Fender2004}.
For X-ray binaries, the thermal emission and the hard X-ray emission are both measured
in X-rays. However, as the thermal emission shifts to the UV energy band for AGN, 
the hardness needs to be defined between the UV and X-ray flux for AGN,
for example as the ratio of the power-law luminosity
against the total luminosity (sum of the power-law and disc luminosity):
\begin{equation}
\label{hardness}
 H =  \frac{L_P}{L_{\rm tot}} = \frac{L_P}{L_P+L_D} .
\end{equation}

The redshift-hardness distribution of the main sample is shown
in Figure~\ref{redshift}. Most of the sources are at the redshift
$z < 1.35$ (see the upper panel of the figure). 
At higher redshifts, 
the UVW1 flux, which is the most commonly used filter in the Optical Monitor,
corresponds to the rest-frame wavelength $\lambda < 1240\AA$.
The sources at higher redshifts are included only if they have significant
measurements in either optical band, U, B, or V.
The highest concentration of the sources is at redshift around $z \approx 1$
and hardness $0 < H < 0.3$.

A mean (median) value of the hardness as a function of redshift 
is shown in Fig.~\ref{redshift} by a blue solid (dashed) line.
The plot indicates that the average value of the hardness is always lower than 0.4.
We note that the typical value of the hardness is significantly lower than a similar
quantity in \citet{Koerding2006a} due to the different calculation
of the disc and power-law luminosities.
Figure~\ref{redshift} also reveals that the average hardness decreases with the redshift.
However, this might be due to an observational bias, when at higher redshift
more luminous sources usually in thermal (soft) states are observed.
The red (dot-dashed) line shows the ratio of the median to mode
(maximum value of the distribution in the given redshift interval).
A large deviation of this curve from the median curve at high redshift indicates
that the maximum value of the hardness at the high-redshift bins is much lower
than for low-redshift intervals.






The radio loudness is defined as the ratio between the radio luminosity
and the total luminosity:
\begin{equation}
{\rm RL} =  \log \frac{L_R}{L_{\rm tot}} = \log\frac{4\pi D_L^2  \nu_{\rm 4.85 GHz} F_{{\rm 4.85 GHz}}}{L_{\rm tot}},
\end{equation}
where 4.85\,GHz corresponds to 6\,cm wavelength. For sources, when the only available radio measurement
is at 20\,cm, the radio flux at 6\,cm is estimated using eq.~\ref{nu}.
For practical purposes, we  deal mainly with a relative radio loudness, 
calculated as the difference between the radio loudness
and the minimal value of the radio loudness of the sample:
\begin{equation}
\label{rl_rel}
{\rm RL}_{\rm rel} = {\rm RL} - {\rm RL}_{\rm min}.
\end{equation}


\section{Results}

\begin{figure*}[tb!]
\includegraphics[width=0.48\textwidth] {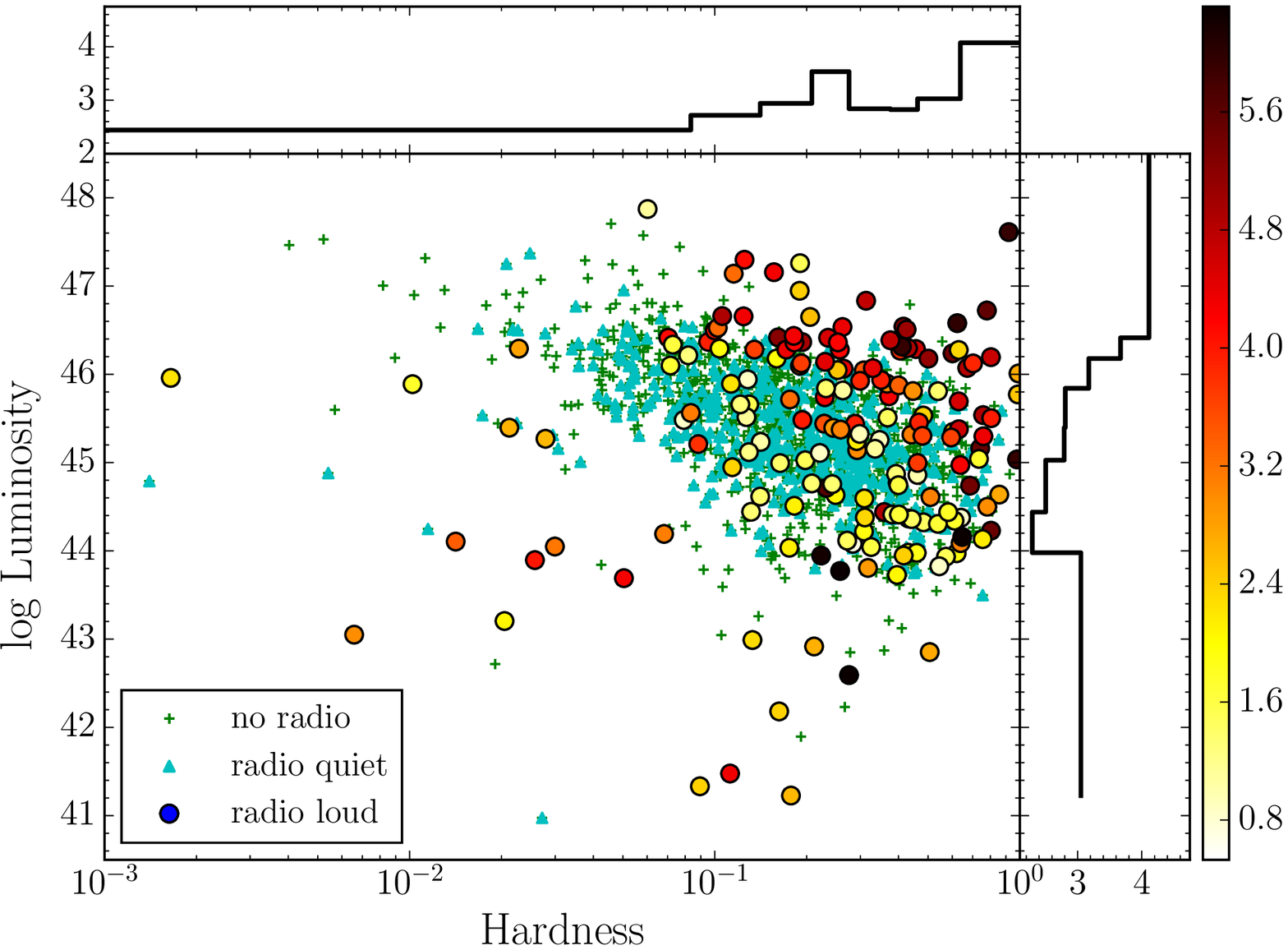}
\includegraphics[width=0.48\textwidth] {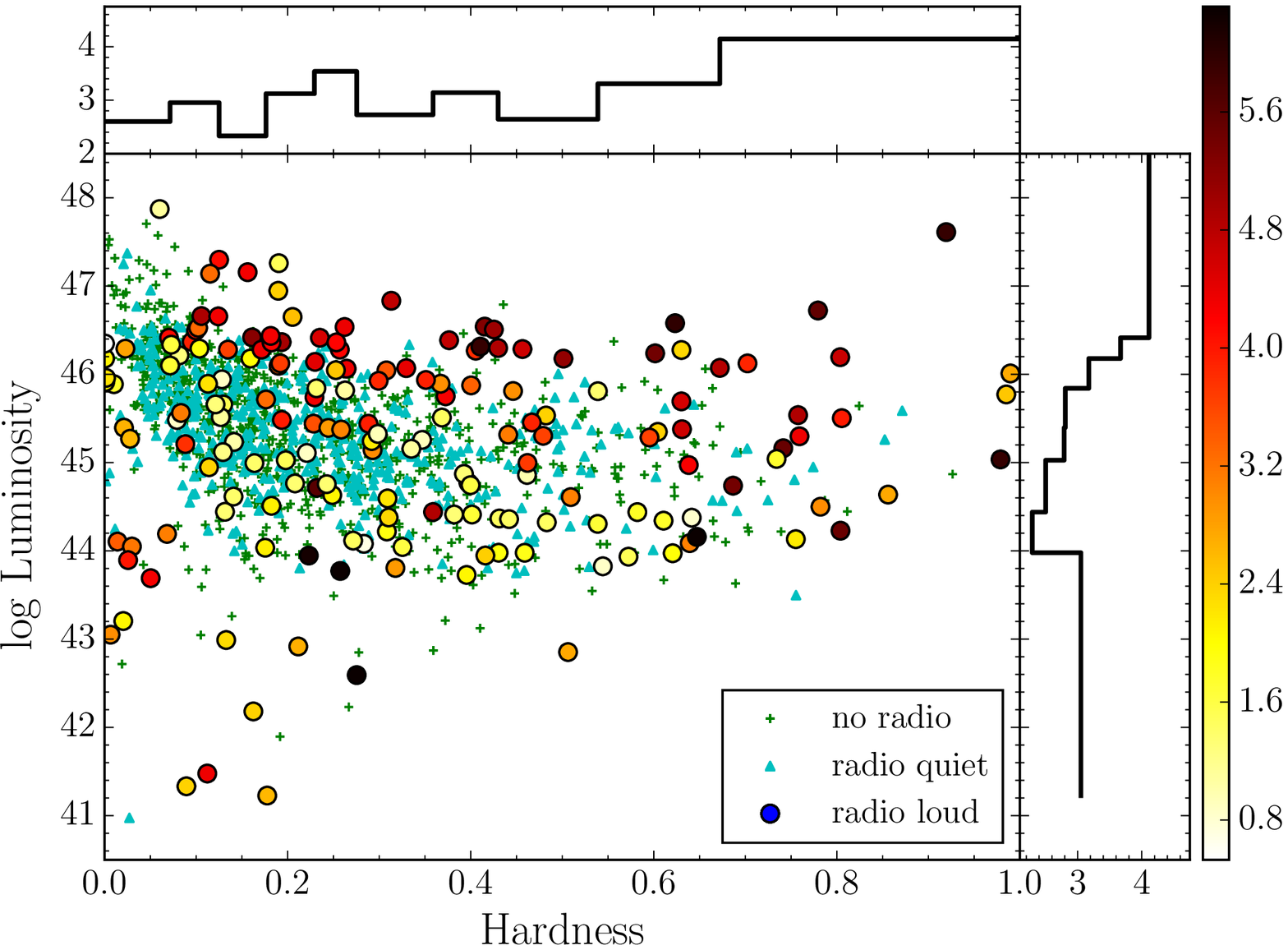}
\caption{
Hardness-luminosity diagram for AGN. 
The coloured circles correspond to sources with 
constrained radio flux measurements, the colour denoting the relative radio loudness of the source.
Radio-quiet sources, which have been detected only with an upper limit 
are shown by cyan triangles.
The smaller green crosses correspond to sources with no available radio flux measurement.
The side histograms show an average value of the relative radio-loudness in respective bins
either in  hardness (top) or luminosity (right).
The histograms are calculated using only sources with constrained radio flux measurements.
The plot is shown in logarithmic scale of  hardness (left) for better comparison
with X-ray binaries and in linear scale (right) to show the hard-state part of the diagram in greater detail.
}
\label{dfld_rmpos}
\end{figure*}



\subsection{Radio-loudness in the hardness-luminosity plane}
\label{hid}

The hardness-luminosity diagram for our main sample is shown in Figure~\ref{dfld_rmpos}.
Each source in the sample is represented by a point in the diagram.
The sources with constrained radio flux measurements
are emphasised by coloured circles whose colour
corresponds to their relative radio loudness.
Sources with no available radio flux measurements are marked by green crosses.
Radio-quiet sources, marked as cyan triangles, are
sources that were observed in radio surveys but only with an upper limit,
which corresponds to the FIRST survey sensitivity limit (about 1mJy).
Such sources have a blank `nR' flag in the V{\'e}ron catalogue.


The plot is shown in two versions using either a logarithmic or a linear hardness scale.
The first (shown in the left panel) is more appropriate for comparison
with X-ray binaries, whose state diagrams are usually shown in  logarithmic scale. 
The second plot shown in  linear scale (right panel) reveals
the hard-state part of the diagram in greater detail.
It is apparent from the plots that the radio-quiet sources are on average
softer with a relatively low hardness, which is less than 0.4 for most sources. 
In contrast, the sources with significant radio emission
are spread over the hardness,
and some of them appear in the upper right corner where radio-quiet
sources are rarely represented.

The side-plot histograms show
an average relative radio-loudness calculated per
 respective bin in hardness (top) and luminosity (right).
Only sources with constrained radio flux measurements are used
to compute these histograms.
To avoid  low-number statistics in the less populated parts of the diagram,
we defined the bins to have an equal number of radio-loud sources in each bin.

The histogram reveals increasing radio loudness with  hardness.
We note that according to the definition of relative radio-loudness in eq.~\ref{rl_rel},
the increment of the average relative radio-loudness by one corresponds
to an increase in the radio luminosity by one order of magnitude.
The histogram of  radio loudness as a function of luminosity 
reveals that the average radio loudness is higher for sources with 
high ($\log{L} > 46$) and low ($\log{L} < 44$) luminosities
(but see how this changes for the Eddington ratio instead of luminosity,
as shown  in Fig.~\ref{dfld_sdss}).

To quantify the statistical significance of the relation between  radio loudness and hardness,
we applied the Pearson statistical test, which gives the Pearson coefficient $r=0.246$
and p-value $p=0.001$. This result suggests a weak but statistically significant positive correlation.
The statistical significance of this correlation is enhanced when low-luminosity sources are not included in the test
(we also show that these sources might have hardness measurements which are systematically
lower than their actual intrinsic values).
For sources with the total luminosity $\log L > 44$, the Pearson coefficient increases to $r=0.322$
and the p-value decreases to $p=5\times10^{-5}$,
indicating that the radio loudness is higher for harder sources and lower for softer sources.

The low-luminosity radio-loud sources might correspond to analogous low-hard states in X-ray binaries.
They standardly populate the lower right corner of the diagram.
However, their position in the AGN diagram, compared to the XRB one, is moved towards the left 
(which is especially apparent in the plot with the linear scale of hardness).
The most likely explanation for this discrepancy is a significant contribution of a host galaxy that is unavoidable in low-luminosity sources
and contributes more to UV than to X-rays.
Consequently, the sources appear much softer in the hardness-luminosity diagram.
In the most extreme case, when their
AGN contribution is negligible in UV
and their entire UV flux is due to the host galaxy, 
such sources would have intrinsically the greatest possible hardness, H=1,
but the `observed' hardness would be much lower.

\subsection{Comparison with non-active galaxies}

\begin{figure}[tb!]
 \includegraphics[width=0.5\textwidth]{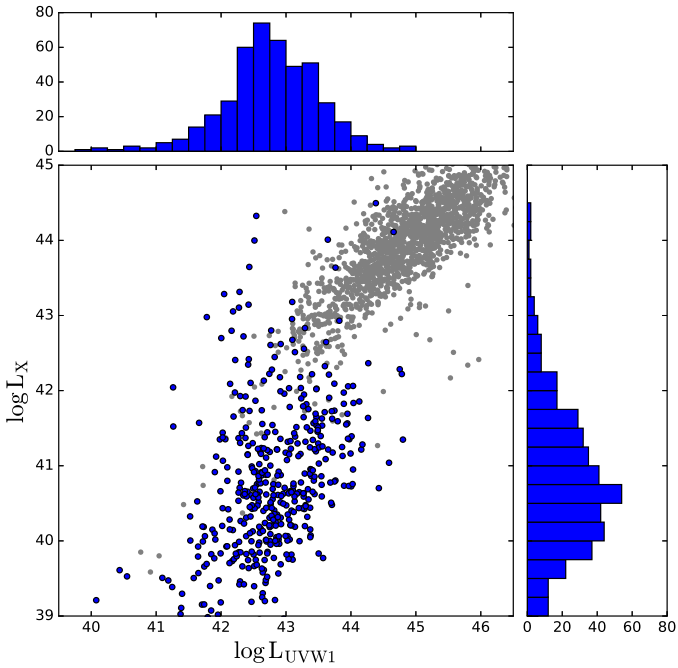}
\caption{Distribution of UV and X-ray fluxes measurements for non-active galaxies (dark blue).
The distribution of AGN is overlaid  (light grey).}
\label{histogram_nonactive}
\end{figure}

\begin{figure}[tb!]
 \includegraphics[width=0.5\textwidth]{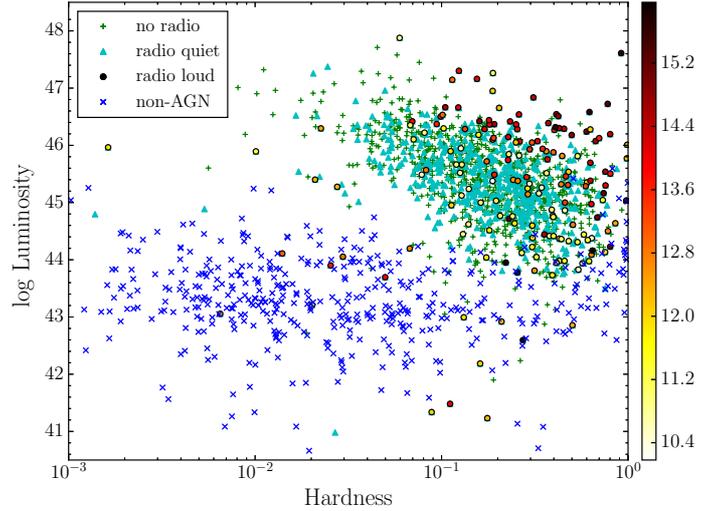}
\caption{Hardness-intensity diagram for non-active (blue crosses)
and active galaxies (the same notation as in Fig.~\ref{dfld_rmpos}).
}
\label{hid_nonactive}
\end{figure}


To better understand the contribution of the host galaxy to the AGN UV spectrum,
we performed a similar analysis to that done with AGN for galaxies
that are considered to be inactive. The sample is created from the simultaneous
{\xmm} observations, as described in Sect.~\ref{samples}, by cross-correlation
with the 2MASS catalogue of galaxies \citep{Huchra2012}. We removed all
sources classified as quasars or AGN in the 2MASS catalogue, 
and also those sources which are present in our AGN sample. We also removed sources
with undetected UV or X-ray (above 2\,keV) flux,
sources with an extended UV emission, and sources with zero or negative redshift.
We applied correction on the Galactic reddening.

The distribution of measured UV and X-ray fluxes for the sample of non-active galaxies
is shown in Figure~\ref{histogram_nonactive}.
The X-ray flux is significantly weaker with respect to the UV flux
when compared to AGN, which are overlaid in the plot and denoted by  light grey dots.
Some of the sources with high X-ray luminosity probably contain
a hidden active nucleus 
because the X-ray luminosity due to starlight in the host galaxy
can hardly exceed $\log L_X \approx 42$ \citep{Gilfanov2014}.
While their X-ray luminosity exceeds the possible contribution
of the host galaxy, the UV luminosity can still be entirely due to the host galaxy,
since the host-galaxy UV contribution peaks at $\log L_{\rm UV, host-galaxy} \lesssim 43$.

Figure~\ref{hid_nonactive} shows the hardness-luminosity diagram 
for both non-active and active galaxies together. 
The disc and power-law luminosity of non-active galaxies are calculated in a similar but simpler way,
as proportional to UV and X-ray luminosity ($L_{\rm disc} \approx 2*L_{\rm UVW1}$,
and $L_{\rm power-law} \approx 10*L_{\rm 2-10 keV}$) for all sources.
The plot is  shown in a logarithmic scale in the hardness for better
clarity.
The plot reveals that the non-active galaxies appear very soft,
which is the consequence of the significantly larger contribution of the 
host galaxy in UV  than in the X-ray domain.

In general, the level of mixing of AGN with non-active galaxies is very low in Fig.~\ref{hid_nonactive}.
However, some non-active galaxies in our sample
exhibit a luminosity exceeding $\log L_{\rm tot} \approx 43$,
where low-luminosity AGN appear.
This implies that the apparent softness of low-luminosity AGN ($\log L < 44$)
may indeed be due to the contribution of the host galaxy.
Therefore, for these sources it is  essential to correctly estimate the host-galaxy contribution.
%
However, the effect due to the host-galaxy contamination
is not easy to estimate from the available data;  a possible treatment
will be discussed in Sect.~{\ref{hg_contamination}}.

\begin{figure}[tb!]
 \includegraphics[width=0.5\textwidth]{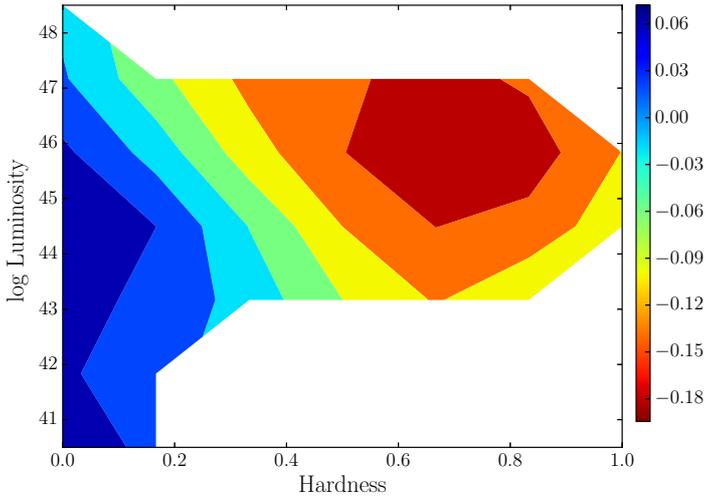}
\caption{Distribution of the photon index difference $\Delta \Gamma = \Gamma - \Gamma_{\rm mean}$ (indicated by colour)
of the whole sample in the hardness-luminosity diagram.
Negative values correspond to harder X-ray spectra and positive values to softer ones.}
\label{dfld_xhardness}
\end{figure}

\begin{figure}[tb!]
 \includegraphics[width=0.5\textwidth]{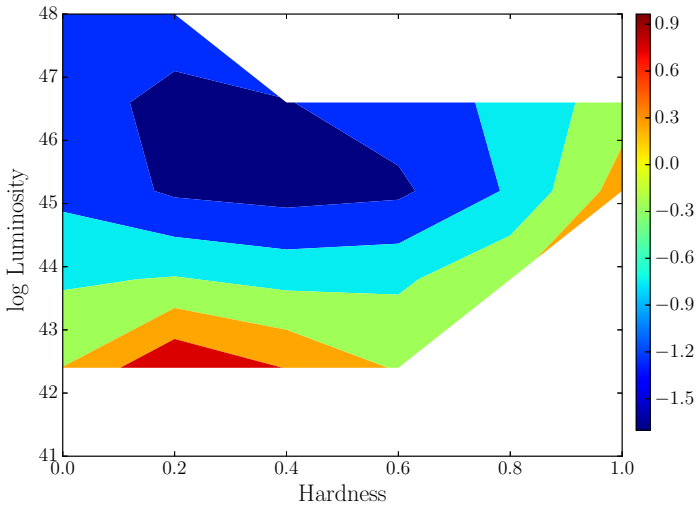}
\caption{Distribution of the UV slope difference $\Delta \beta = \beta - \beta_{\rm mean}$ (indicated by colour) in the hardness-luminosity diagram . 
Only sources with measured UV slope from the XMM-Newton
Optical Monitor are considered.
}
\label{dfld_omslope}
\end{figure}

\subsection{UV and X-ray spectral slope in the hardness-luminosity plane}

Higher sensitivity of the X-ray and UV measurements
by XMM-Newton allows us to construct hardness-luminosity diagrams,
which -- instead of  radio loudness --
show the X-ray hardness and UV slope, respectively.
To have a more detailed look at the hard-state part of the diagram,
we use the linear scale of the hardness in these plots. 
Because there are  a relatively large number of sources with measured X-ray and UV slopes, 
we show these plots using coloured contours instead of scatter plots.
Figure~\ref{dfld_xhardness} shows the relative difference of the X-ray photon index
with respect to its average value:
$\Delta \Gamma = \Gamma - \Gamma_{\rm mean}$. 
Negative values of $\Delta \Gamma$ (red) correspond to lower values of the photon index
and thus flatter X-ray spectra. The locations of these sources  in the diagram
are consistent with the expected location of sources in the hard state.


Similarly, Figure~\ref{dfld_omslope} shows 
the relative difference in the UV slope, defined as 
$\Delta \beta = \beta - \beta_{\rm mean}$.
Only sources with measured UV slopes are used for this plot.
The UV slope is calculated in the wavelength domain
and thus the positive value (red) corresponds
to a flatter spectrum.
Steeper spectra are characteristic of the disc thermal emission,
while flat UV spectra indicate that the UV emission might be dominated
by other spectral components. 
Flat UV spectra are especially found at low-luminosity sources, which might indicate
a large effect of the host galaxy contribution there.

\begin{figure*}[tb!]
 \includegraphics[width=0.49\textwidth]{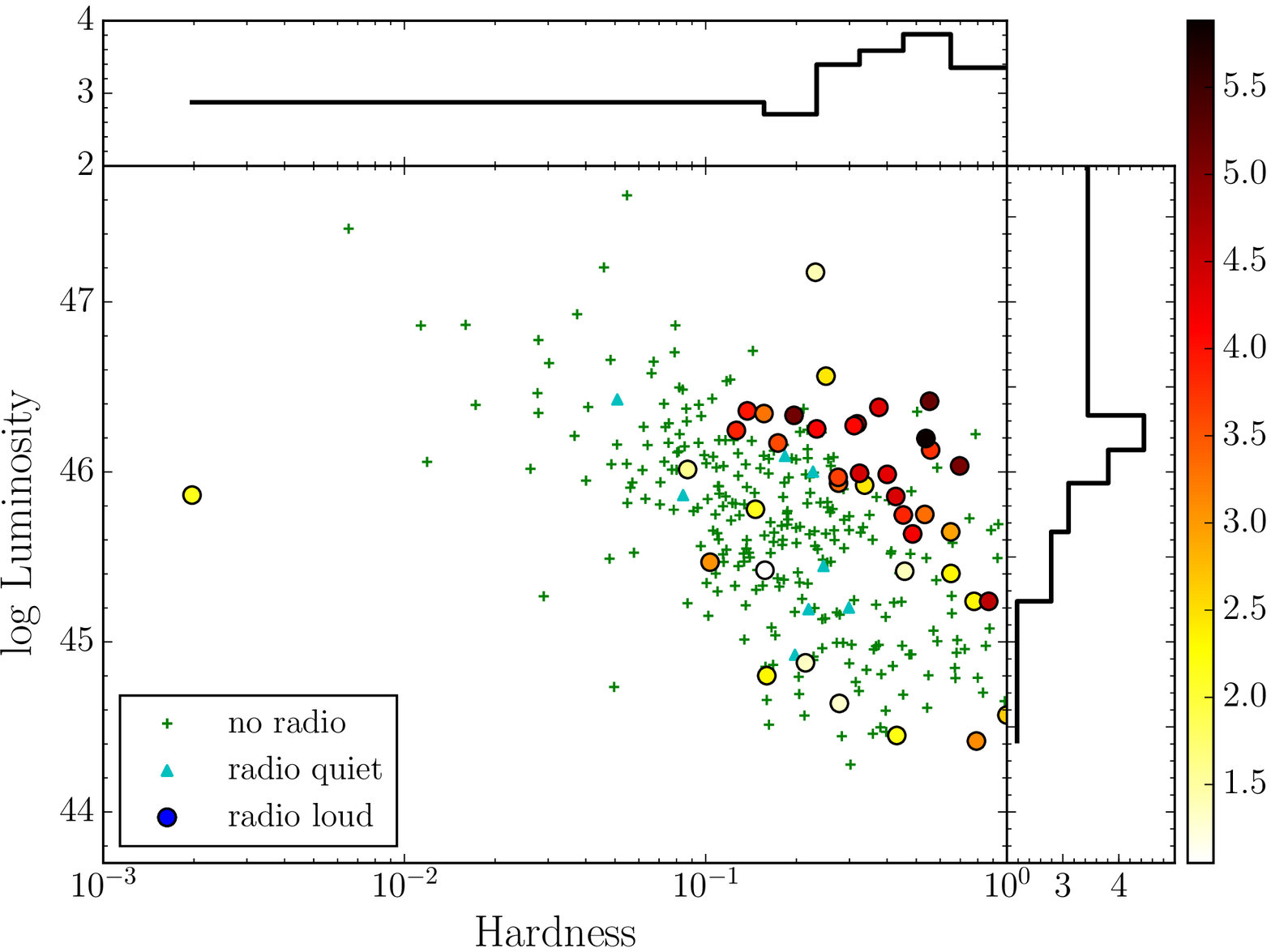}
 \includegraphics[width=0.49\textwidth]{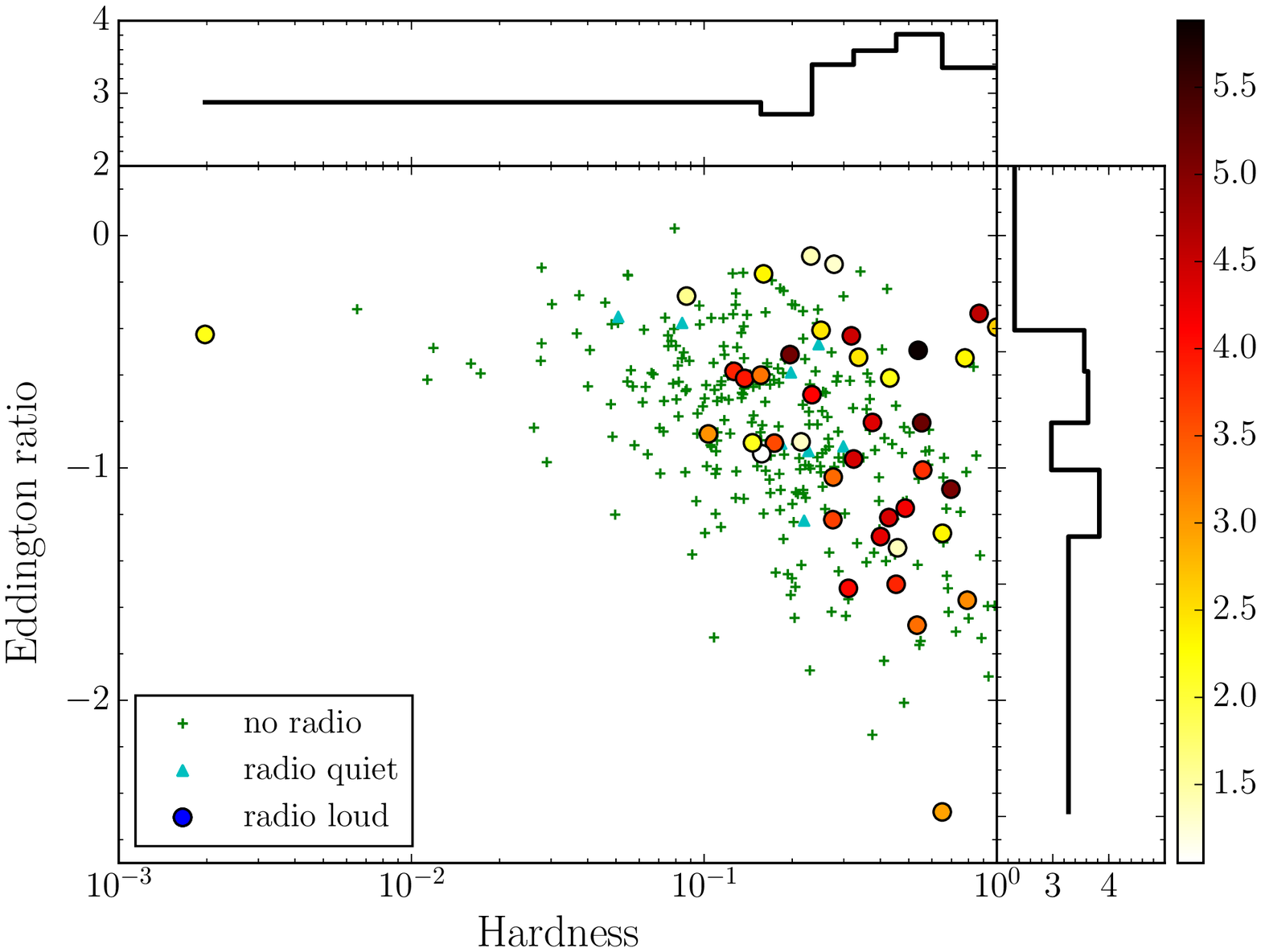}
\caption{
Radio-loudness in the hardness-luminosity (left)
and the hardness-Eddington ratio (right) diagram for 
sources of the XMM-SDSS subsample.
}
\label{dfld_sdss}
\end{figure*}

\subsection{XMM-SDSS subsample}


A significant part of our sample has complementary measurements in the optical
domain from the Sloan Digital Sky Survey (SDSS; \citet{Gunn2006}).
For sources from the 7th SDSS data release,
\citet{Shen2011} also constrained
 the black hole mass, which may be used to scale the luminosity to the Eddington luminosity.
As the Eddington ratio better corresponds to the real accretion state than the total luminosity,
especially given that our sample spans a wide range in the black hole mass,
we decided to use only the 316 sources sources with measured black hole mass.
We note that the mass estimates are the virial masses,
which may be discrepant from the true black hole masses (see \citet{Shen2011} 
for a more detailed discussion), but they are the best available
estimates we have for our analysis (see also Sect.~\ref{discussion_mass}).

Fig.~\ref{dfld_sdss} shows the radio-loudness in the hardness versus luminosity plot (left panel)
and versus the Eddington ratio (right panel) for the XMM-SDSS subsample.
Both plots show a higher concentration of radio-loud sources in the harder part of the diagram.
However, the vertical positions of several sources significantly change between the plots.
The luminosity plot shows that the radio-loud sources are on  average more luminous
than the radio-quiet sources. On the other hand,
the distribution of the Eddington ratios between radio-quiet and radio-loud sources is comparable.
The average radio-loudness is stronger for lower-accreting sources,
while it decreases when the accretion rate approaches the Eddington limit.

Sources at the low Eddington ratio are clearly harder than most sources in the sample. 
An average value of the hardness is $H_{\rm SDSS, mean} \approx 0.25$ for the whole subsample,
while for the low Eddington ratio sources (${\lambda_{\rm Edd}<-1.4}$) the hardness is $H_{\rm low Edd.} \approx 0.5$.
The XMM-SDSS subsample does not contain any sources whose luminosity
is lower than $\log L_{\rm tot} \approx 44$, and thus the effect of a host galaxy
is not prominent in this sample.
The diagram with the Eddington ratio is
more reminiscent of the analogous diagram for X-ray binaries,
indicating that the Eddington ratio is a more appropriate proxy of the accretion rate.



\section{Discussion}
\label{discussion}

\subsection{Similarity of the AGN and XRB spectral-state evolutionary diagrams}

The AGN hardness-luminosity diagrams (Figs.~\ref{dfld_rmpos} and \ref{dfld_sdss})
can be qualitatively compared with an analogous diagram created for a sample of X-ray binaries 
(e.g. Fig.~8 of \citet{Dunn2010}).
The global appearance is quite similar for sources above $\log L_{\rm tot} > 44$.
Radio-loud sources concentrated towards the right part of the hardness-luminosity diagram
indicate that jets are launched when the sources are in the hard states
characterised by a higher ratio between the X-ray and UV emission, harder X-ray spectra,
and flatter UV spectra.
In contrast, the jets seem to be almost absent in the soft states (upper left part of the state diagrams),
when the sources are characterised by a much stronger UV bump due to the thermal disc emission
and softer X-ray spectral slope,
confirming the long-held belief that most luminous AGN are indeed radio quiet.

A remarkable difference between our AGN (Fig.~\ref{dfld_rmpos}) and XRB diagrams
was found in the low-luminosity range. 
Low-luminosity AGN are softer than the low-hard state analogies would be in the hardness-luminosity diagram. 
The most likely explanation for this difference
is a significant contribution of the host galaxy to the UV flux.
This `galaxy dilution' problem is well known in surveys 
selecting AGN via their optical/UV emission \citep[see e.g.][]{Merloni2016}, 
and is the most severe limitation when attempting to characterise 
genuine analogues of the low-hard state in AGN samples.
A proper decomposition of the nuclear and host-galaxy contribution is therefore crucial
to correctly place the low-luminosity sources in the diagram, 
and we discuss this in the following section.




%
%

\begin{figure}[tb!]
\includegraphics[width=0.5\textwidth]{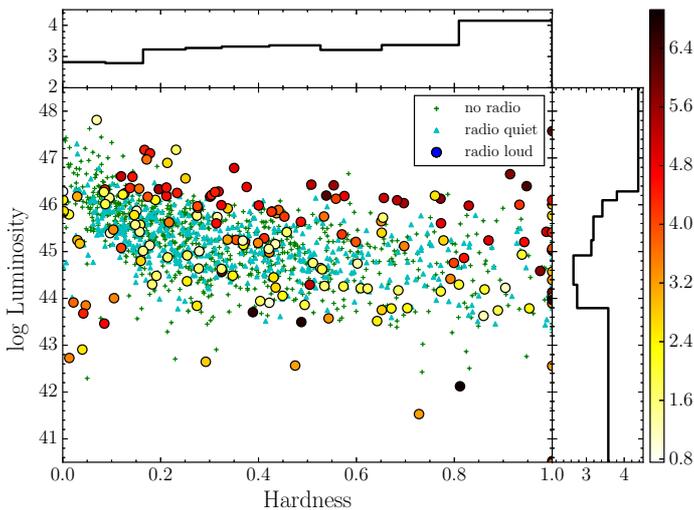}
\caption{Hardness-luminosity diagram
with applied host-galaxy correction according to  eqs.~\ref{eq_uvsfr} and \ref{eq_xsfr}.
The notation is the same as in Fig.~\ref{dfld_rmpos}. 
The side histograms show an average value of the relative radio-loudness in respective bins
either in  hardness (top) or luminosity (right).
}
\label{dfld_HG}
\end{figure}

\subsection{Host galaxy contamination}
\label{hg_contamination}
%
%


In the absence of very high spatial resolution observations, available only for the nearest galaxies,
the only reliable deconvolution of the host galaxy
contamination is obtained by a detailed modelling of the AGN spectral energy distribution (SED)
\citep[see e.g.][and references therein]{Ho1999, Ho2008}.
This has been done so far for only nearby low-luminosity AGN,
and there is not yet a fully accepted consensus on the UV emission
of a special class of low-luminosity AGN, the LINERs.
While \citet{Ho2008} concluded that these sources completely lack the UV emission from the nucleus,
\citet{Maoz2007} found a significant contribution that can be associated with the nuclear activity.

In the first case, where all UV flux is from the star light, 
the affected sources should be moved entirely to the right 
in the diagram (the hardness would be equal to one).
For the second case, we can compare the results by \citet{Maoz2007}
with the measurements of our final sample.
However, the only source from his study that fulfils 
our strict filtering criteria for creating the final sample is NGC 3998.
Its hardness from the analysis by \citet{Maoz2007}
is $H_{\rm Maoz,\, NGC\,3998} = 0.86$.
The hardness from our main sample without any host-galaxy correction
is  $H_{\rm NGC\,3998} = 0.21 \pm 0.01$,
which is significantly lower.
This clearly indicates that the disc luminosity is largely overestimated in such sources.

A detailed analysis of the host-galaxy contribution of all low-luminosity sources
in our sample is beyond the scope of this paper.
However, we could use some empirical relations to estimate its contribution.
One possibility would be to use the mean values of the UV and X-ray flux measurements of non-active galaxies
shown in Figure~\ref{histogram_nonactive}. 
However, the AGN hosts may be even brighter
since the nuclear activity may be related to the enhanced star formation,
as shown by, e.g. \citet{VanDenBerk2005} and \citet{Santini2012}. 
Subtracting an average host-galaxy contribution
based on non-active galaxies might  therefore be insufficient.

A tight correlation between the emission of the host galaxy and bolometric
AGN luminosity was found by \citet{Netzer2009}. Therefore, we used the relation
obtained from their Fig.\,13 to estimate the star-formation luminosity as
\begin{equation}
 L_{\rm SFR} \approx 2 \times 10^8 \, L_{\rm bol}^{0.8},
\end{equation}
where $L_{\rm SFR} = \nu L_{\rm {\nu}, 60\,{\mu}m}$ is the infrared luminosity measured at 60 micrometres, 
and $L_{\rm bol}$ is the bolometric luminosity.

We used the total luminosity $L_{\rm tot}$ ($ = L_{\rm disc}$ + $L_{\rm power-law}$)
as an approximate estimate of the bolometric luminosity.
We further constrain the star formation rate ${\rm (SFR)}$ using the relation by \citet{Calzetti2010}\footnote{
$L_{\rm SFR}$ is constrained from the infrared luminosity $\nu L_{\rm {\nu}, 70\,{\mu}m}$
at 70\,${\mu}m$ in the relation by \citet{Calzetti2010}, but we made an assumption that 
within the general uncertainties we can assume
that the infrared spectrum between these two bands is relatively flat \citep[see e.g.][]{Symeonidis2016}.}:
\begin{equation}
 {\rm SFR} = 0.6 \times \frac{L_{\rm SFR}}{10^{43}}.
\end{equation}
The UV luminosity can be then estimated from the SFR
using a relation 
based on GALEX NUV data \citep{Murphy2011, Hao2011, Kennicutt2012}:
\begin{equation}
\label{eq_uvsfr}
  \log(\nu L_{\nu, \rm UV, SFR}) = \log({\rm SFR}) +  43.17.
\end{equation}
Because the average frequency of the UVW1 and GALEX-NUV bands differ,
we multiplied the obtained UV luminosity by the ratio between these frequencies 
(i.e. by a factor $2301/2910 \approx 0.79$) to estimate the contribution of the host galaxy in $L_{\rm UVW1}$.

Similarly, we can estimate the X-ray luminosity from the SFR \citep{Gilfanov2014, Mineo2014}.
We used the relation by \citet{Gilfanov2014}:
\begin{equation}
\label{eq_xsfr}
 L_{\rm 0.5-8\,keV} = 2.5\times10^{39} SFR.
\end{equation}

%
Figure~\ref{dfld_HG} shows the hardness-luminosity diagram after the applied host-galaxy correction
using eqs.~\ref{eq_uvsfr} and \ref{eq_xsfr}.
The plot is shown in the linear scale for hardness to focus on changes in the hard part of the diagram,
and can  therefore be directly compared to the right panel of Fig.~\ref{dfld_rmpos}. 
The radio-loud sources  moved to the right in the diagram,
towards higher hardness as expected.
For NGC 3998,
we get a `corrected' value of the hardness $H_{\rm NGC\,3998} = 0.48 \pm 0.01$,
which is, however, still insufficient.

We note that 
while the detailed spectral decomposition would be the best
way to treat the low-luminosity sources,
several indirect methods can be employed in future
extensions of current work.
One way to estimate the intrinsic UV emission
of low-luminosity sources is from the broad H-alpha luminosity \citep{Stern2012},
i.e. from the reprocessing of the central flux in the broad-line region.
Intrinsic luminosities can also be estimated from the reprocessed emission 
at the torus, which is observed in the mid-IR band. This emission is closely
related to the intrinsic emission from the central AGN \citep{Asmus2015}.
However, the spectral energy decomposition of a large sample of AGN
with deep multi-wavelength observations is clearly the most reliable way.

\subsection{Effect of the intrinsic reddening on measured UV slopes}
\label{discussion_extinction} 

Not only hardness might be affected by the host galaxy, but also the measured UV slope.
The UV slope shown in the binned AGN hardness-luminosity diagram (Figure~\ref{dfld_omslope})
was calculated according to eq.~\ref{beta} and was corrected for the Galactic
extinction. However, it was not corrected for an intrinsic extinction due to the host galaxy.

The intrinsic extinction could be estimated  as in eq.~\ref{Galactic_extinction}.
However, the ratio of the total to selective extinction $R_V$  typically
has a different shape than the Galactic profile \citep[see e.g.][]{Calzetti1994, Richards2003, Gaskell2007}.
Based on the SDSS quasar extinction measurements by \citet{Richards2003},
\citet{Czerny2004} derived a simple empirical formula,
\begin{equation}
 \frac{A_{\lambda}}{E(B-V)} = -1.36 + 13 \log\left(\frac{1}{\lambda}\right) \, \left(\rm {\mu}m\right),
\end{equation}
valid in the range of $(1/{\lambda})$ between 1.5 and 8.5  $\left(\rm {\mu}m\right)^{-1}$.

We used this formula to estimate extinction at UVW1 (2910 $\AA$), UVM2 (2310 $\AA$), and UVW2 (2120 $\AA$) wavelengths.
We get $A_{\rm 2910 \AA} = 5.6$, $A_{\rm 2310 \AA} = 6.9$, and $A_{\rm 2120 \AA} = 7.4$, respectively.
For the difference in the measured and intrinsic UV slope between the UVW1 and UVM2 bands,
we obtained
\begin{equation}
 \Delta \beta_{\rm UVW1-UVM2} \approx 5.185*E(B-V).
\end{equation}
The difference in the UV slope is thus linearly proportional to the intrinsic reddening.
For high intrinsic reddening, $E(B-V) \gtrsim 0.1$,
the UV slope difference between the intrinsic and measured value is $\Delta\beta_{\rm UVW1-UVM2}  \gtrsim 0.5$.
However, for a typical intrinsic reddening of quasars, \citet{Gaskell2004} concluded that more than
90 \%\ of the SDSS quasar sample has intrinsic reddening $E(B-V) < 0.055$ \citep[see also][]{Vasudevan2007, Lusso2010}.
This corresponds to $\Delta\beta_{\rm UVW1-UVM2} \approx 0.25$, which 
is smaller than the distribution of uncorrected UV slopes (see Fig.~\ref{dfld_omslope}). 
However, the intrinsic reddening might be significantly higher in the case of low-luminosity sources,
for which a more robust analysis is needed to estimate the effect of the intrinsic reddening,
and this should be investigated in future work.

%


\begin{figure}[tb!]
 \includegraphics[width=0.5\textwidth]{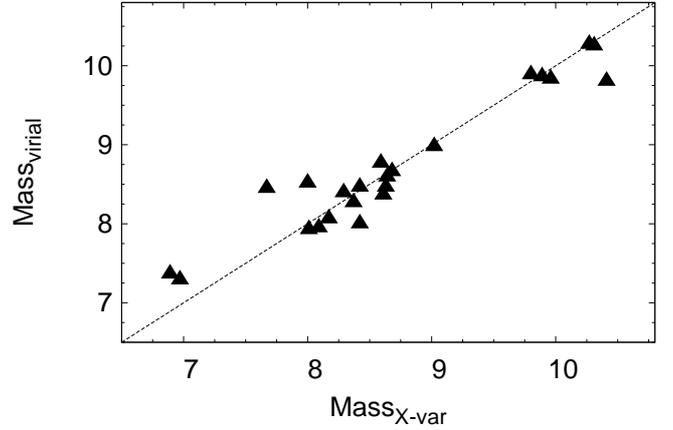}
\caption{Comparison of the mass measurements from the X-ray variability
by \citet{Ponti2012} with the virial mass constraints from \citet{Shen2011}.
The dashed line represents the 1:1 ratio, showing that the  methods
are in  very good agreement.
}
\label{mass_ponti_shen}
\end{figure}

\subsection{Mass and hardness-ratio (Eddington) diagrams}
\label{discussion_mass}

A significant uncertainty in the hardness-luminosity diagrams for AGN
is the mass of the central black hole.
The range of possible masses is large, in extremes from $10^5$ to $10^{10}$
solar masses. Therefore, a more appropriate quantity
than the luminosity is the luminosity
scaled by the black hole mass, i.e. the Eddington ratio. 
The most reliable method for determining mass, the reverberation technique \citep{Peterson2004},
has been applied only to a small sample of local AGN.
Therefore, the black hole mass in AGN is 
usually constrained from empirical correlations 
between the mass and the luminosity of a specific emission line
or the continuum flux density at a specific wavelength in the optical band; however, it is affected by a large uncertainty 
\citep[see e.g.][]{Denney2009}.

Another possible technique to constrain the black hole mass is through X-ray 
variability \citep{Zhou2010,Ponti2012}. 
Although some information about the X-ray variability
is reported in the most recent release of the {\txmm} catalogue,
there are several reasons why this information cannot be directly used. 
First, the reported values
correspond to maximal variability detected over the exposure,
and so the real variability is,  in general, lower.
Second, the variability measurements
are sensitive to high background-flare events that need to be omitted
from the light curve used for the calculations. 
Last, we found that the reported values are often not consistent between different detectors (PN and MOS).
Therefore, the light curves would need to be properly analysed in detail,
which is beyond the scope of this paper.

Instead, we used the virial black hole mass measurements from the SDSS catalogue
\citep{Shen2011}. The estimates are done on the width of H$\alpha$,
H$\beta$, Mg II, and C IV emission lines. Although these measurements
might be affected by systematic uncertainties
(unavoidable in the conversion between the line luminosity and the mass
or bolometric luminosity), we found that there is a quite good agreement
of measured masses for sources with the mass constrained 
from the X-ray variability (see Fig.~\ref{mass_ponti_shen}).

\begin{figure}[tb!]
 \includegraphics[width=0.49\textwidth]{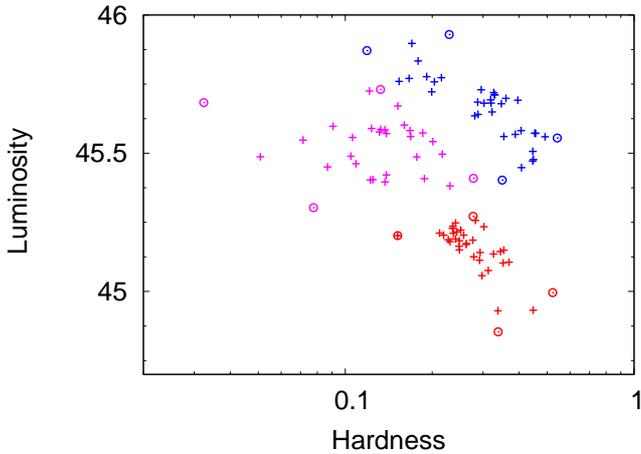}
\caption{
Hardness-luminosity diagram for three sources with the largest number 
of observations available in the {\xmm} catalogue:
CDFS J03321-2747 (red), CDFS J03324-2741B (blue), CDFS J03324-2740 (magenta).
The crosses represent the simultaneous data, while the open circles
show the extreme values of any possible combinations of UV and X-ray fluxes.
}
\label{simul_three_sources}
\end{figure}

\begin{figure}[tb!]
 \includegraphics[width=0.49\textwidth]{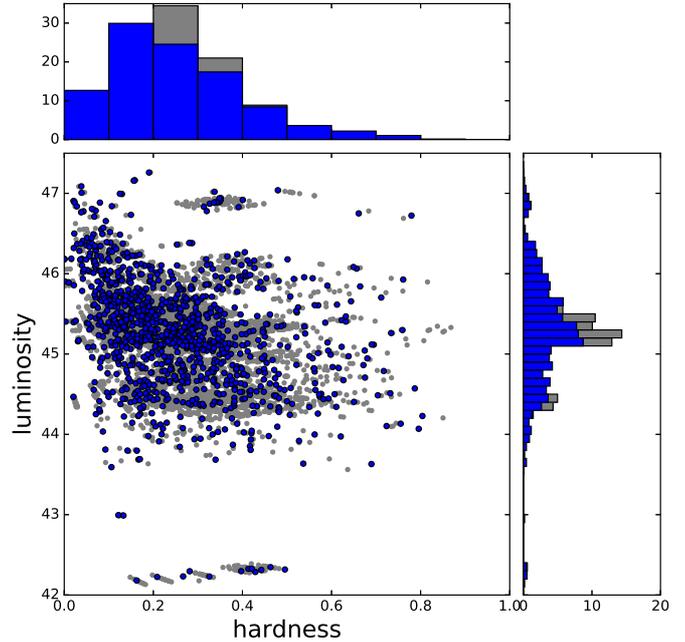}
\caption{
Hardness-luminosity diagram for sources with multiple observations available in the {\xmm} catalogue.
The blue points represent all simultaneous UV and X-ray observations,
while the grey points represent all possible combinations of measured UV and X-ray flux for each source with multiple observations.
The side-plot histograms show fractional distribution (in percent) of the hardness and luminosity for simultaneous measurements (blue boxes)
and all possible combined measurements (grey boxes), respectively.
}
\label{simul_all_sources}
\end{figure}

\subsection{Variability of source positions in the hardness-luminosity diagrams and effects of non-simultaneity}

Thanks to the simultaneity of the UV and X-ray observations by {\xmm},
 in our analysis we can rule out any effects of source variability
when observations from different epochs are compared.
However, the UV and X-ray emission are not supposed to originate at exactly the same region.
While the X-rays are believed to come from a more central region (X-ray corona, X-ray reflection at the innermost accretion disc),
the UV flux may be dominated by emission coming from more distant parts of the accretion disc.
Thus, despite the simultaneity of observations, some variability of source positions 
in the hardness-luminosity diagrams can still be expected in both hardness and luminosity. 
Therefore, we further investigate the variability of source positions in the diagram
for sources with multiple observations available.

Figure~\ref{simul_three_sources} shows the hardness-luminosity diagram
for three sources with the highest number of different observations in the {\xmm} catalogue.
The hardness and luminosity constrained from the simultaneously measured UV and X-ray fluxes
are shown by crosses, while the open circles show the extreme values obtained from any possible
combination of measured UV and X-ray flux without requiring its simultaneity.
The lowest hardness is obtained when the minimum X-ray and maximum UV flux are combined,
and vice versa for the highest hardness.
Similarly for luminosity, the highest value is obtained by a combination of maximum values
of measured fluxes and the lowest from minimum ones.
The plot shows a quite large scatter in the hardness-luminosity diagram position even for simultaneous data.
The extreme values obtained by a combination of (non-simultaneously) measured minimum and maximum values
of UV and X-ray fluxes are not largely offset from the simultaneous data.
This clearly illustrates that the scatter in the hardness-luminosity relation is not
dominated by the non-simultaneity of the data, but rather by some intrinsic variability of the accretion-disc corona system.

The luminosity-hardness plane 
for all sources with available multiple observations in the {\xmm} catalogue
is shown in Figure~\ref{simul_all_sources}.
The simultaneous measurements (blue points) are overlaid 
on   all possible combinations of their non-simultaneous measurements (grey points).
The side-plot histograms show the fractional distribution of different values of hardness and luminosity.
Because the distributions are not very different,
a strict simultaneity of UV and X-ray measurements is not critical in this kind of analysis.
This is consistent with a similar conclusion made by \citet{Vagnetti2010}, who analysed
$\alpha_{ox}$ distribution of the AGN observations by {\xmm}.
They also realised that the intrinsic variability is significantly greater
than any artificial variability caused by non-simultaneity.

\subsection{Sample selection and biases}

We have shown the AGN state diagrams for a sample of {bona fide}
identified AGN in the {\xmm} EPIC and OM serendipitous source catalogues.
We excluded blazars, whose X-ray emission is dominated by 
a relativistically boosted radiation from collimated jets,
extended sources,  and also 
heavily X-ray obscured sources where no direct X-ray emission
from the nuclear region is expected to be detected.
The advantage of the sample
is that it contains a large number of sources sensitively and simultaneously
measured in the UV and X-ray bands by the same satellite.

Still, the sample is not homogeneous enough
to be directly used to investigate the source density
in different parts of the diagram and compare it with
X-ray binaries, for which this quantity was constrained by \citet{Dunn2010}.
%
Deep and wide surveys are necessary
to have a more homogeneous sample.
The main advantage of such surveys
is the availability of multi-wavelength measurements, which allow  a more detailed analysis,
in particular to properly calculate the UV de-reddening and contribution of the host galaxies.
This is most critical for low-luminosity sources, which are, however, not well
represented in the existing samples. Therefore, even deeper and more sensitive surveys
are needed to significantly improve the analysis done in this paper.

The next X-ray missions, such as e-ROSITA \citep{Predehl2014} or Athena \citep{Nandra2013},
 will provide  data with the required sensitivity, as will the planned radio facilities, such as 
ASKAP/EMU in the southern hemisphere or APERITIF/WODAN in the north 
\citep[for a review see e.g.][]{Norris2013},
and finally the SKA radio interferometer \citep{Dewdney2009}
will significantly increase the sensitivity of the radio measurements.

\section{Conclusions}
\label{conclusions}

We studied the analogy between the accreting black holes in X-ray binaries
and active galactic nuclei by investigating the spectral states of AGN 
and their relation to the radio properties.
We employed simultaneous UV and X-ray observations of AGN
performed by the {\xmm} satellite in the period 2000-13
to compare their total luminosity with  hardness, 
defined as the ratio of the X-ray power-law luminosity 
against the total luminosity composed of thermal disc
emission, as well as the non-thermal corona X-ray emission.

We found that  radio loudness increases on average with  hardness,
suggesting that the jet emission occurs in the hard accretion states,
consistently with X-ray binaries and with previous results
obtained by \citet{Koerding2006a} using non-simultaneous measurements of quasars.
By analysis of sources with multiple observations, we realised that 
the scatter in the hardness-luminosity relation is not dominated 
by the non-simultaneity of the data, 
but rather by some intrinsic variability of the accretion-disc corona system. 

The high sensitivity of the detectors on board the {\xmm} satellite also allowed us to
study  low-luminosity sources ($\log L_{\rm tot} < 44$). 
However, we found that especially their UV emission is strongly affected
by their host-galaxy contamination, leading to a significant underestimate of the hardness.
A more detailed spectral decomposition is needed to properly constrain
the intrinsic luminosity of such sources.



We showed the AGN state diagrams for a subsample of sources
that were present in the SDSS survey
and for which virial black hole mass estimates were available.
The Eddington ratio calculated from the black hole mass
is more appropriate for characterising the accretion state than the total luminosity.
The hardness-Eddington ratio diagram revealed that
the average radio-loudness is higher for low-accreting sources,
while it decreases when the accretion rate is close to the Eddington limit.

%


A significant improvement in our analysis can be achieved with
more sensitive multi-wavelength observational surveys. The three most important aspects are the
precise measurements of the radio, UV, and X-ray intrinsic luminosities
properly corrected for the host-galaxy contamination;
the large size and homogeneity of the sample; and the reliable mass measurements
of central supermassive black holes.

\begin{acknowledgements} 
The authors thank the anonymous referee for the helpful suggestions and comments
that led to a significant improvement of this work.
JS also acknowledges very useful discussions with Ignacio de la Calle,
Giovanni Miniutti, Sara Elisa Motta, Margherita Giustini, Gabriele Ponti, Barbara de Marco, Marion Cadolle Bel, Erin Kara, and Victoria Grinberg.
This research was financially supported from the Grant Agency of the Czech Republic
within the project No.\,14-20970P.
The research was also partially funded by 
the European Union Seventh Framework Program (FP7/2007–2013) under grant 312789.
The presented work is based on observations obtained with XMM-Newton, 
an ESA science mission with instruments and contributions 
directly funded by ESA Member States and NASA.
\end{acknowledgements}

\bibliographystyle{aa} 
\bibliography{/home/jirka/Documents/references} 

\section*{Appendix A1. Details of the main sample creation}
 \label{appendix-1}
\setcounter{table}{0}
\renewcommand{\thetable}{A\arabic{table}}

\begin{table}[b!]

\caption{Filtering procedure to create the main sample.}
\centering
\begin{tabular}[width=0.5\textwidth]{cc}
 \rule{0cm}{0.5cm}
 Filtering & Number of sources \\
 \hline \hline
 \rule[-0.7em]{0pt}{2em} 
 Total number of AGN entries$^{a}$ & 6188 \\
  \rule[-0.7em]{0pt}{2em} 
 Removing sources with extended UV emission & 5407 \\
 \rule[-0.7em]{0pt}{2em} 
 Removing X-ray underexposed sources & 4204 \\
 \rule[-0.7em]{0pt}{2em} 
 Removing sources with $\Gamma < 1.5$ or $\Gamma > 3.5$  & 3085 \\
 \rule[-0.7em]{0pt}{2em} 
 Removing sources with their measured UV flux &\\
 corresponding to $\lambda \lesssim 1240\,{\AA}$ in their rest frame  & 2336 \\
 \rule[-0.7em]{0pt}{2em} 
 Selecting the best observation for each source & 1525 \\
 \rule[-0.7em]{0pt}{2em} 
 Excluding nuclear HII regions & 1522 \\
 & \\
\end{tabular}
\tablefoot{$^{a}$\,Obtained by a cross-correlation of the XMM-Newton catalogues with the   \citet{Veron-Cetty2010},
SDSS, and XMM-Cosmos catalogues.
}

\label{app_final_sample} 
\end{table}

In this section, we describe the details of the procedure used to create our main sample.
The main parent samples are the third XMM-Newton
serendipitous source catalogue, data release 6 \citep{Rosen2016},
which contains 9160 observations with more than  half a million  clear X-ray detections,
and the XMM-Newton serendipitous ultraviolet source survey \citep{page2012},
which includes 7170 observations of more than 4.3 million different sources.
Requiring the quasi-simultaneity by linking the identification
observation numbers, we found 49,842 matched entries between the {\xmm} UV and X-ray catalogues.

From this parent sample, we selected only AGN using the V\'{e}ron  \citep{Veron-Cetty2010},
SDSS \citep{Alam2015}, and XMM-COSMOS \citep{Hasinger2007} survey catalogues as described in Section~2.
For the SDSS AGN subclass,
we used the redshift measurements that were obtained from the fits
without using quasar templates, denoted as `z\_noqso' \citep{Bolton2012}.
We selected only sources with reliable redshift measurements
(i.e. with `zwarning\_noqso = 0').

When there was positive cross-matching from multiple catalogues,
we preferentially adopted values for the redshift measurements in this order:
1) SDSS (redshift from visual inspection of the spectra), 2) the Veron catalogue, and 3) XMM-COSMOS. However, if the value of the  redshift
was less than or equal to zero in a more preferred catalogue, we chose the value
from the other catalogue.
We only considered sources with redshift higher than $z_{\rm min} = 0.001$.

Table~\ref{app_final_sample} summarises how the sample was further slimmed down 
by applying various filtering to avoid uncertainties related with extended UV emission,
low significance of detections, etc.

\end{document}